\RequirePackage{fix-cm}
\documentclass[twocolumn]{svjour3}          
\usepackage[square, comma,numbers, sort&compress]{natbib}
\smartqed  
\usepackage{graphicx}
\usepackage{float}
\usepackage{overpic}
\usepackage{pict2e}
\usepackage{siunitx}
\usepackage{amsmath}
\usepackage{supertabular}
\usepackage{subfigure}
\usepackage{multirow}
\usepackage{hyperref}					%

\usepackage{xcolor}

\newcommand{\name}[1]{#1}

\DeclareMathOperator{\sech}{sech}
\begin{document}

\title{The In-plane Expansion of Fractured Thermally Pre-stressed Glass Panes
}

\subtitle{An equivalent temperature difference model for engineering glass design
}

\author{Jens H. Nielsen \and Michael A. Kraus \and Jens Schneider  
}

\institute
            {Assoc. Prof. Dr. J. H. Nielsen \at
             Technical University of Denmark\\
         \email{jhn@byg.dtu.dk}           
           \and
           Prof. Dr. J. Schneider \at
              Technical University of Darmstadt\\ 
             \email{schneider@ismd.tu-darmstadt.de}   
      \and
          Dr. M. A. Kraus \at 
          ETH Z\"urich\\ 
             \email: {kraus@ibk.baug.ethz.ch}
}
\date{Received: date / Accepted: date}

\maketitle
\begin{abstract}
The present paper is concerned with deriving simplified design equations and charts for modelling in-plane expansion of fractured thermally pre-stressed glass panes using the method of \emph{equivalent temperature differences} (ETD) together with a thermal expansion analogy for strains. The starting point is a theoretical method based on linear elastic fracture mechanics merged with approaches from stochastic geometry to predict the 2D-macro-scale fragmentation of glass. The approach is based on two influencing parameters of glass: (i) fragment particle size, $\delta$, and (ii) fracture particle intensity, $\lambda$, which are related to the pre-stress induced strain energy density, $U_\text{D}$, before fracture. Further Finite Element (FE) analysis of single cylindrical glass particles allow for establishing functional relations of the glass fragment particle dimensions, the pre-stress level and the resulting maximum in-plane deformation. Combining the two parts of two-parameter fracture pattern modelling and FE results on fragment expansion, formulas and engineering design charts for quantifying the in-plane expansion of thermally pre-stressed glass panes due to fracturing via an ETD approach is derived and provided within this paper. Two examples from engineering practice serve as demonstrators on how to use our ETD approach to compute the equivalent temperature difference and resulting internal forces as well as deformations. This approach serves furthermore as a basis to estimate secondary effects (such as fracture-expansion-induced deformations or stresses) on support structures or remaining parts of glass laminates in form of handy ETD load cases within analytical as well as FE analysis.

\end{abstract}



\section{Introduction and Motivation}
The use of glass as a structural material in engineering requires the analysis and prediction of its behaviour in the intact state, during the fracture process and, after the fracture process is completed, in the post-fracture limit state. Thermally tempered glass is often used as a structural material due to its superior strength properties compared to annealed glass. Due to the large amount of residual stresses present, tempered glass typically fractures into small dices that are less harmful than shards from broken annealed glass. It is therefore also called safety glass if the number of fragments per unit area is large enough. Contrary, if used for laminated glass, the post-fracture behaviour of tempered glass is often considered unfavourable due to the small fragments that lead to a global membrane-like structural behaviour (where almost no bending stiffness is left) of the laminated glass plates in case all glass plies are fractured. Another interesting effect is the rapid in-plane expansion of the tempered glass fragments during the fracturing process. The fracture front travels at about \SI{1500}{m/s} \cite{Nielsen2009} and a significant part of the strain energy is released during this fracturing process. Within the process, each fragment expands, eventually resulting in a significant global expansion of the fractured glass plate. This expansion can lead to (i) flying debris in monolithic (single layer) glass plates, (ii) to global bending in glass laminates, where only one or several plies of a multi-laminate set-up fracture, and (iii) to delamination of the interlayer or in-plane failure of the broken glass if the expansion is strongly impeded by the remaining plies. To account for this effect in engineering glass design and to provide a handy, yet realistic, estimation of this effect, a simplified model for computing the increase in size of a glass pane made of thermally pre-stressed glass after fracture was developed within this paper. A clear analogy to thermal expansion of continuum materials given an equivalent temperature difference (ETD) is hereby followed.

\section{Theoretical Background and State of the Art}
This section lays the foundation for the deduction of our engineering approximation model of in-plane expansion of fractured pre-stressed glass panes. 


\subsection{Background on Thermally Pre-Stressed Glass}
\begin{figure} 
\includegraphics[width=\columnwidth]{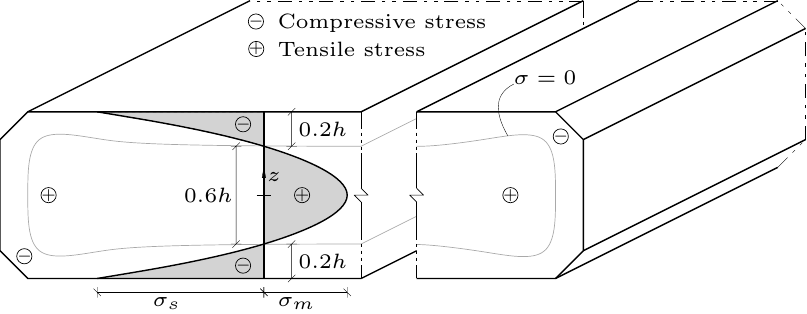}
\caption{Stress distribution in the far field area of a tempered glass plate and a sketch showing a contour line at zero stress (dotted line)}
\label{fig:1}
\end{figure}

The strength of standard float glass is governed by its tensile strength, which itself is significantly influenced by small flaws in the surface, reducing the typical engineering tensile strength of annealed float glass to somewhere in the interval \SI{30}{\MPa} to {100}{MPa}, see e.g. \cite{Schneider2016a}. Inducing a residual stress state by thermal tempering, a greater resistance to external loads, together with a certain fracture pattern in case of failure can be obtained. This leads to a desired level of safety with respect to human injuries due to the smaller fragments. For these reasons, thermally pre-stressed glass is also known as tempered safety glass. The residual stress obtained during the tempering process is characterised by its approximately parabolic distribution through the thickness, where compressive stresses reside on the outer surfaces, which is balanced by internal tensile stresses in the mid-plane, cf. Figure~\ref{fig:1}. This parabolic stress distribution, $ \sigma(z) $, can be written in terms of the surface stress, $ \sigma_\text{s} $, and glass thickness, $h$, as:
\begin{equation} \centering
\sigma(z)=\frac{1}{2}\sigma_\text{s}(1-3\zeta^2),\qquad\zeta=\frac{2z}{h}
\label{eq:sig_z1}
\end{equation}

using symbols defined in Figure~\ref{fig:1}. The parabolic stress distribution is in equilibrium and symmetric about the mid-plane. The magnitude of the surface stress is approximately twice the tensile stress, $ (2\sigma_\text{m}=-\sigma_\text{s}) $. The zero stress level is at a depth of approximately $ 21\% $ of the thickness, $ h $, from the surface, known as the compressive zone depth \cite{Nielsen2021}. The surface flaws then are in a permanent state of compression by the compressive residual stress at the surface, which has to be exceeded by externally imposed stresses due to loadings before failure and fracture of the glass pane can occur \cite{Schneider2016a,Nielsen2010b}. The magnitude of residual stresses depends on processing- and material parameters and is not within the scope of this paper. For a deeper insight on this, the reader is referred to other literature such as \cite{Gardon1965,Schneider2016a,Nielsen2010c,Nielsen2014,Nielsen2010a}.

\begin{figure}
	\subfigure{\includegraphics[width=0.241\textwidth]{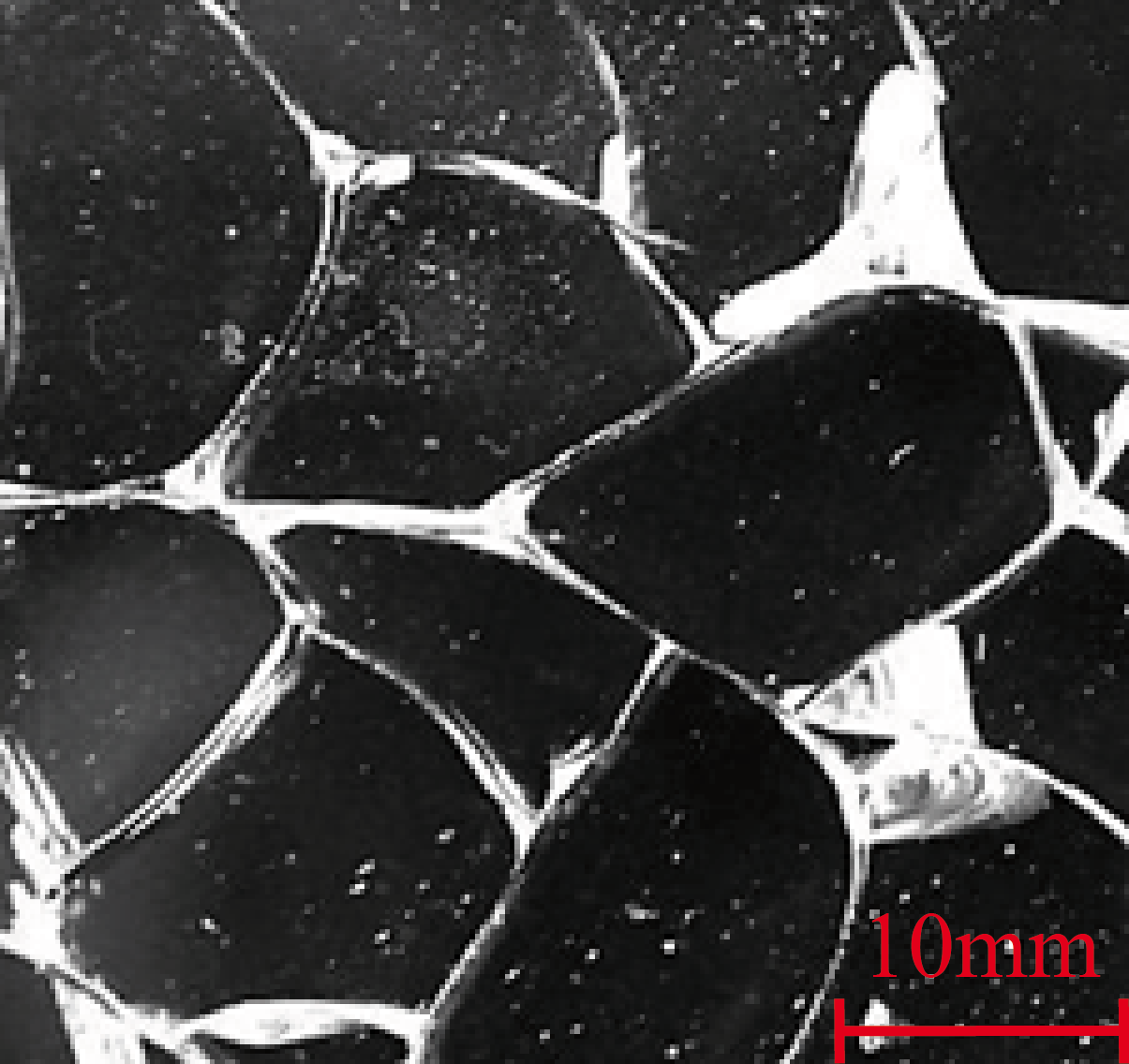}}
	\subfigure{\includegraphics[width=0.23\textwidth]{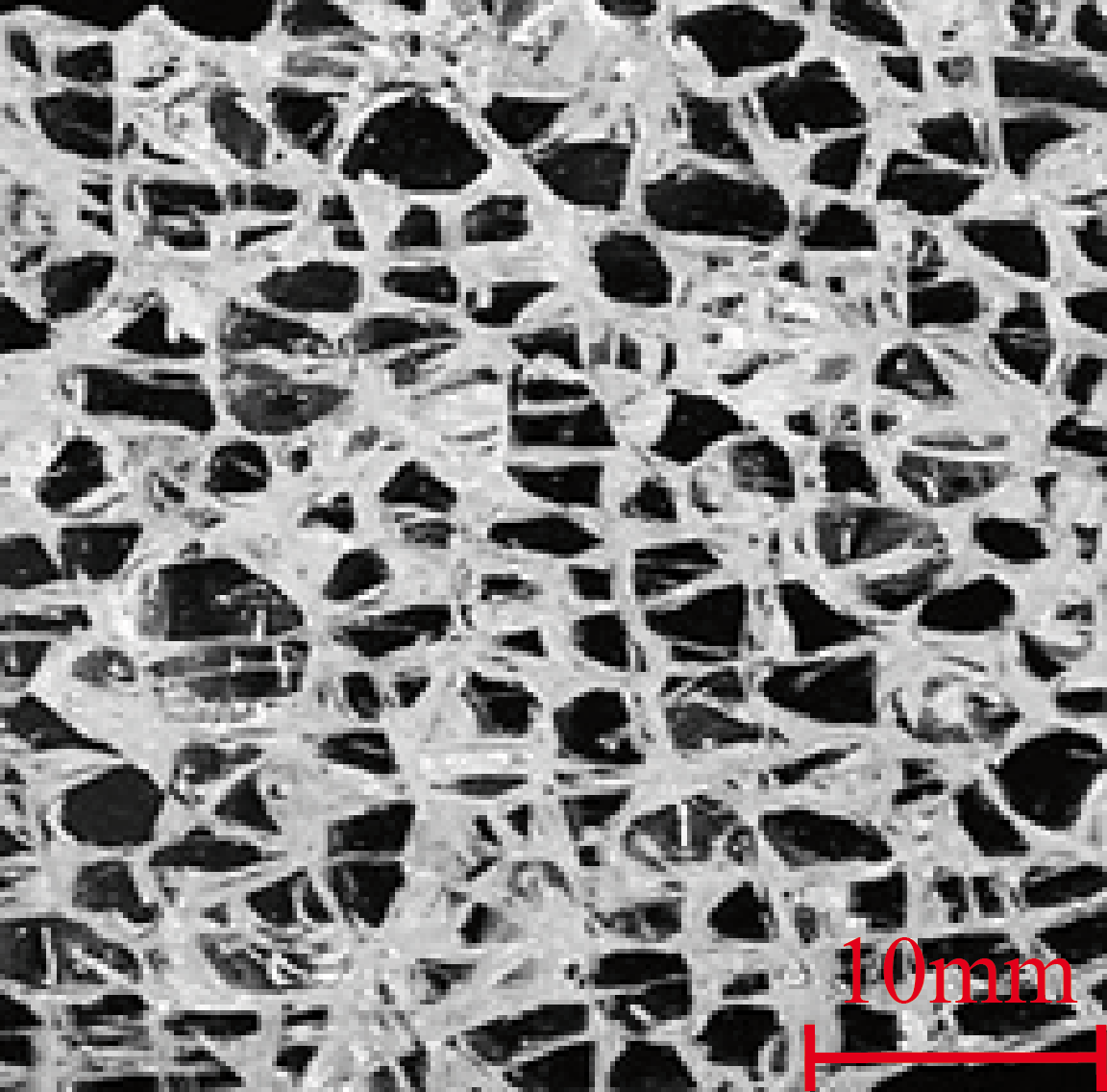}}
\caption{Fragmentation of glass plates with the same thickness of 12 mm for (left) low stored strain energy ($U = 78.1~J/m^2$), (right) high stored strain energy ($U = 354.7~J/m^2$), from \cite{Pourmoghaddam2019a}}
\label{fig:2}
\end{figure}

\subsection{Fracture of Thermally Pre-Stressed Glasses and Glass Fracture Pattern Modelling} \label{subsec:FractStats}
For the case of fracture, thermally pre-stressed glass panes fragmentize completely into many pieces, if the equilibrated residual stress state within the glass plate is disturbed sufficiently and it holds an elastic strain energy large enough. A commonly used example demonstrating this are the so-called "Prince Rupert's drops", possessing bulbous heads and thin tails. These glass drops can withstand high impact or pressure applied to the head, but ''explode'' immediately into small particles if the tail is broken, see e.g. \cite{Hooke1665,Silverman2012,Kooij2021}. The fragmentation is the direct consequence of the release of elastic strain energy stored inside the material due to the residual stress state. The fragment size depends on the amount of the released energy. Small fragments are caused by a high energy release such as the high residual stress state found in tempered glass originating from the quenching process. Lower residual stress states result in larger fragments due to lower stored strain energy (cf. Figure~\ref{fig:2}). Thus, not only the stress but also the thickness of the glass plate plays a role in determining the strain energy. The strain energy density, $U_\text{D}$, which is the strain energy, $U$, normalised with the thickness. The strain energy density then becomes a thickness independent quantity for characterising the energy state of a thermally pre-stressed glass pane, and can then be derived to yield \cite{Pourmoghaddam2019a}:
\begin{equation} \centering
U_\text{D} =\frac{U}{h}= \frac{1}{5} \frac{(1-\nu)}{E} \sigma_\text{s}^2 = \frac{4}{5} \frac{(1-\nu)}{E} \sigma_\text{m}^2 
\label{eq:U_04}
\end{equation}
where $U_\text{D}$ is the amount of elastic strain energy stored in the system per unit volume and thus only depends on the residual stress and the material properties. The above equation is fully in line with other derivations of the strain energy in tempered glass as provided by e.g. \cite{Nielsen2016,Barsom1968,Gulati1997,Warren2001,Reich2012}.

\begin{figure} 
\includegraphics[width=0.45\textwidth]{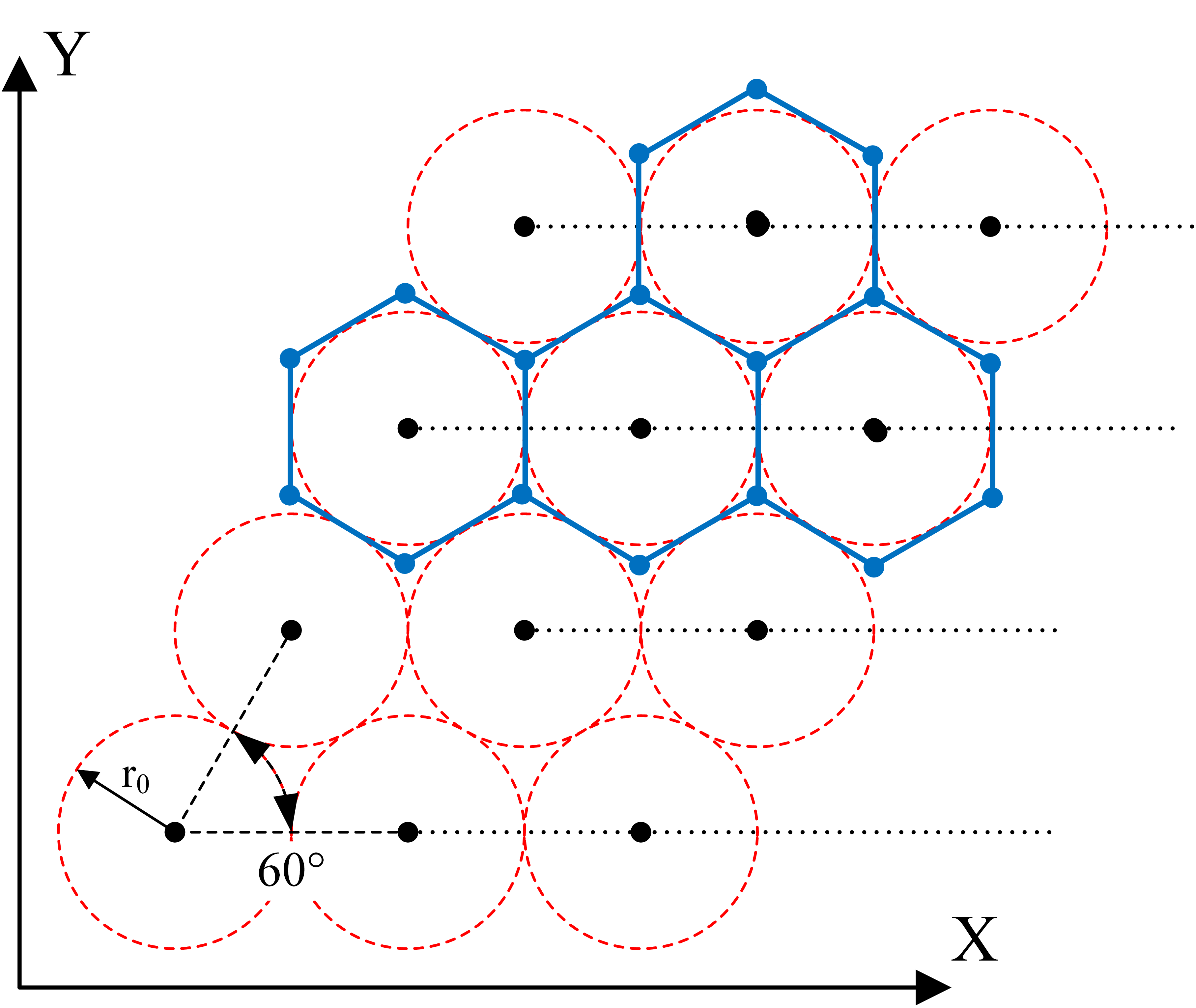}
\caption{Hexagonal close packing (HCP) distribution of points and resulting honeycomb pattern due to the Voronoi tessellation of HCP distributed seed points, from \cite{Pourmoghaddam2019a}}
\label{fig:4}
\end{figure}

\begin{figure} 
\includegraphics[width=0.49\textwidth]{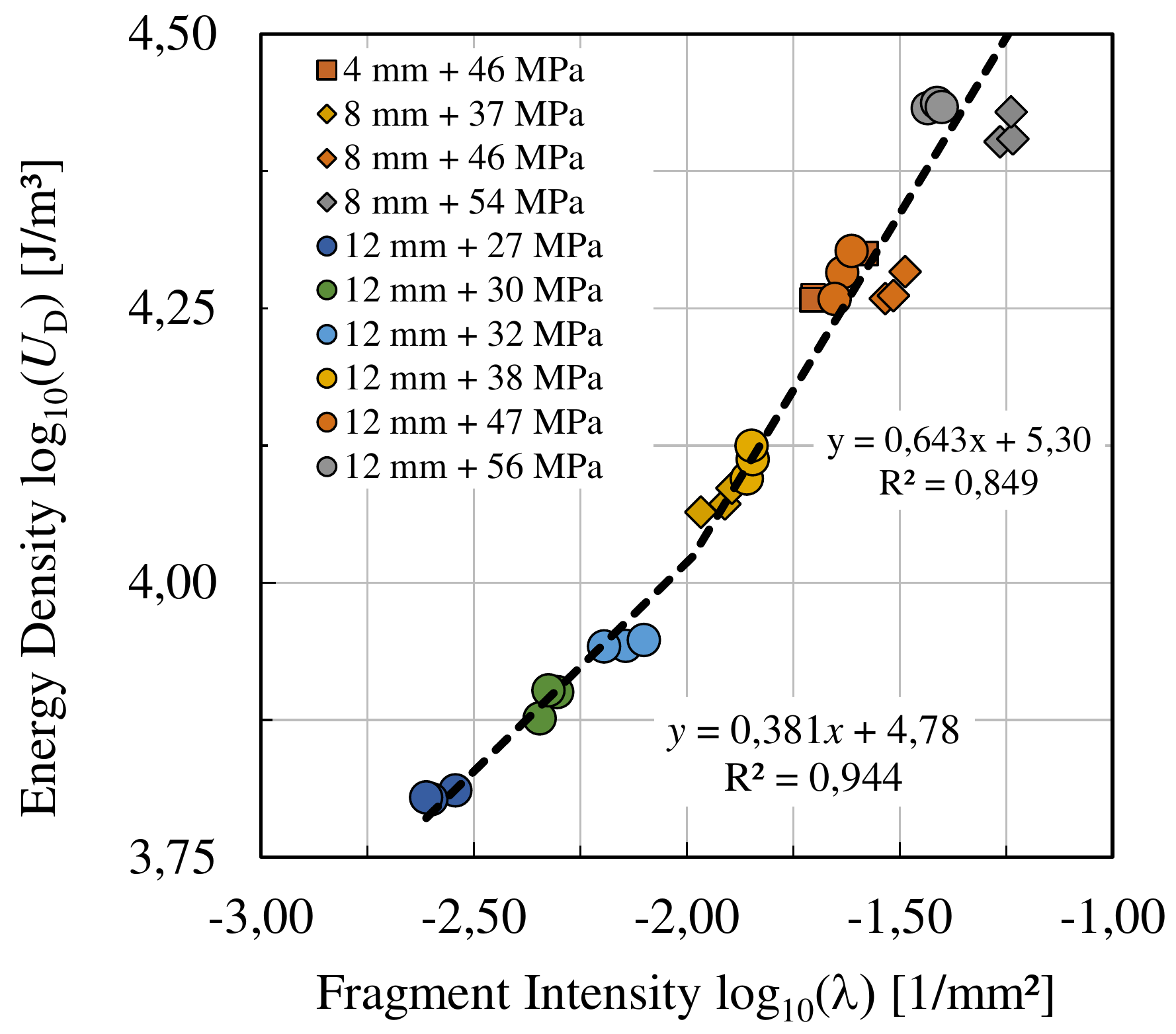}
\caption{Plot of the intensity $\log_{10}(\lambda)$ - strain energy density $\log_{10}(U_\text{D})$ relation, marking two scales for the fracture pattern}
\label{fig:Ud_lambda}
\end{figure}

The papers \cite{Pourmoghaddam2018,pourmoghaddam2018geometrical,pourmoghaddam2018prediction,Pourmoghaddam2019a,kraus2019break} discuss the properties of fragment size, $\delta$, fragmentation intensity, $\lambda$, and the tessellation pattern that result from the fragmentation process, where especially in \cite{kraus2019break,kraus2019PhDThesis} a statistical evaluation and Bayesian treatment of the fracture pattern model is presented. The fracture pattern model combines an energy criterion of linear elastic fracture mechanics and tessellations induced by random point patterns. Statistical analysis of the glass fracture patterns of a comprehensive experimental programme, consisting of thermally pre-stressed glass panes with different thicknesses and levels of thermal pre-stress, allowed for a sound and exhaustive investigation the fracture pattern of tempered glass in order to determine characteristics of the fragmentation pattern (e.g fragment size, $\delta$, fracture intensity, $\lambda$, etc.). The basic modelling approach consists of the idea, that the final fracture pattern is a Voronoi tessellation induced by a stochastic point process, whose parameters can be inferred by statistical evaluation of pictures of several fractured glass specimen. By calibration of a stochastic point process and consecutive tessellation of the region of interest, statistically identically distributed realisations of fracture patterns of a glass pane can be generated. The evaluations there quintessentially show, that the size, shape and number of fragments strongly and non-linearly depend on the strain energy density, $U_\text{D}$, (cf. Figure~\ref{fig:Ud_lambda}) and the fracture pattern may be approximated by a \name{Voronoi}-tessellation induced by a \name{Mat\'ern-Hardcore}-Point-Process. The literature, \cite{pourmoghaddam2018geometrical,kraus2019PhDThesis} show, that the fracture pattern is on average a Hexagonal Close Packing (HCP) with the uniform distance $\delta_\text{HCP}$ between any two adjacent nuclei. This is caused by the dynamic fracture properties as derived by \cite{Yoffe1951} and experimentally verified for tempered glass in \cite{Nielsen2009}. Thus the mean fracture pattern of thermally pre-stressed glass panes is a regular honeycomb with hexagonal cells, cf. Figure~\ref{fig:4}. 

The spread in fragment sizes is often found to follow the power law size distributions \cite{grady2008fragment,Kooij2021}, having only dimensionless fit parameters and contain no characteristic length scale, i.e. they are scale invariant. However, \cite{Kooij2021} can show, that unstressed glass plates follow a hierarchical breakup process with power law size distribution while stressed glass plates follow a random (Poisson) process with fragments showing an exponential size distribution with a natural characteristic length as a fit parameter linked to the residual stress. \citeauthor{Kooij2021} \cite{Kooij2021} found, that the characteristic length scale of the exponential size distribution is approximately the thickness of the plate, $h$, which is in agreement with the findings of \cite{kraus2019PhDThesis}.

In cf. Figure~\ref{fig:Ud_lambda} the variables of the experiments from \cite{Pourmoghaddam2018} "pre-stress" and "thickness" are graphically encoded by colour and symbols respectively. Our analysis delivers two patterns for the relationship between energy density $U_\text{D}$ and the fragment intensity $\lambda$ (both in $\log_{10}$-scale). It is especially interesting, that the patterns are associated with two scales:

\begin{flalign}
\log_{10}(U_\text{D}) = \begin{cases}
0.381 \cdot \log_{10}(\lambda) + 4.78 & \text{if } \lambda \in 10^{[-3;-2]} \\ 
0.643 \cdot \log_{10}(\lambda) + 5.30 & \text{if } \lambda \in 10^{[-2;-1]} 
\end{cases} 
\end{flalign}
\label{eq:lgo_Ud_lambda}
\label{eq:U_lambda_eq}

Despite that novel finding of two separated functions for relating fracture pattern parameters and strain energy density, $U_D$, the expansion model derived in Section~\ref{sec_ETD} will be based on the more simple relation reported in \cite{Pourmoghaddam2019a} for the relation between the fragment size parameter $\delta = 2r_0$ and the strain energy density $U_\text{D}$ is used:
\begin{equation}\label{eq_r_0_simple}
    U_\text{D} = \frac{\SI{122.1}{\joule\per\meter\squared}}{\delta} \quad \Leftrightarrow \quad r_0 = \frac{\SI{61.05}{\joule\per\meter\squared}}{U_\text{D}}.
\end{equation}

\begin{figure}[ht!] 
\includegraphics[width=0.46\textwidth]{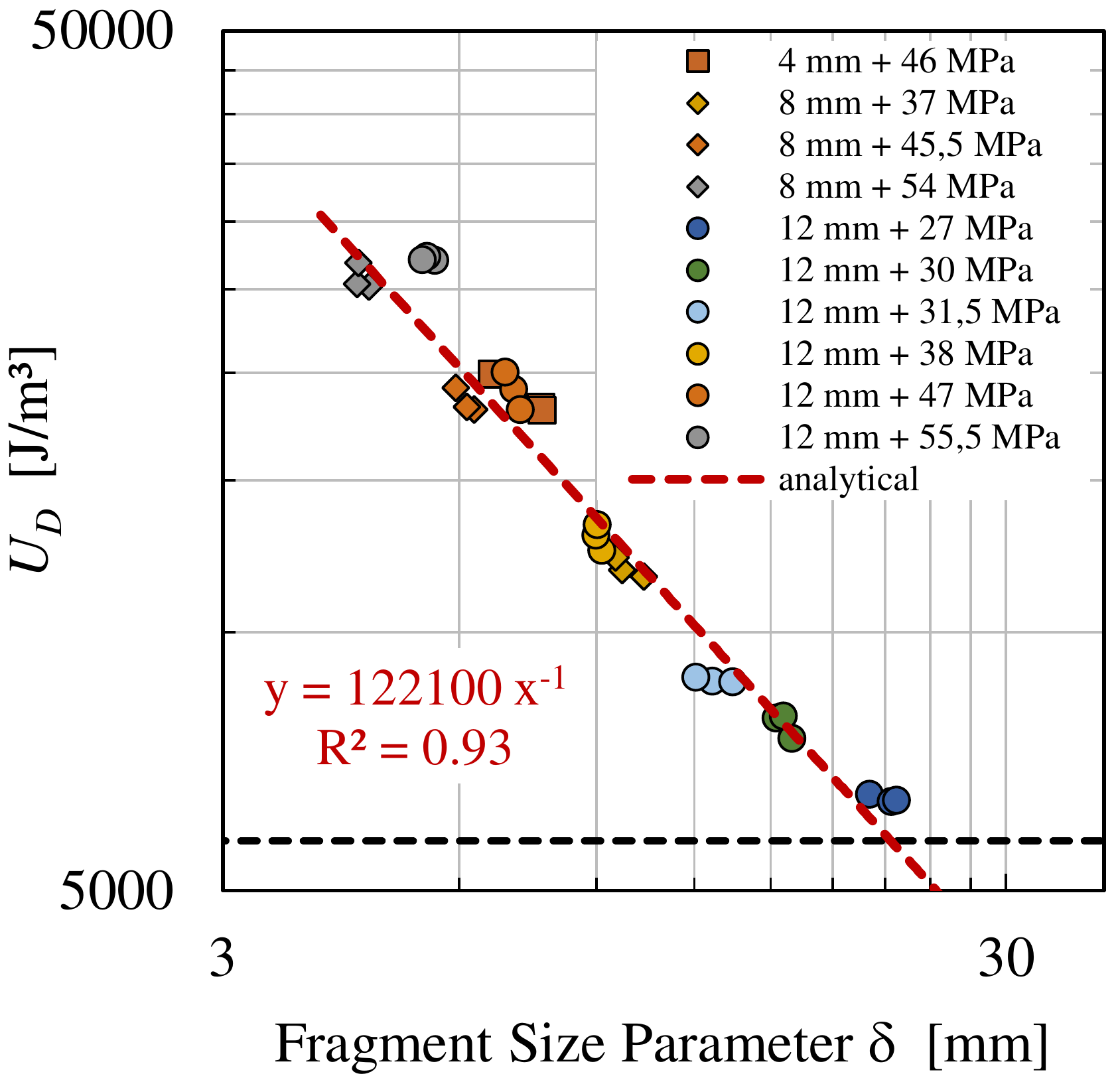}
\caption{Elastic strain energy density, $U_\text{D}$, versus Fragment size parameter, $\delta$. Adapted and enhanced using data from \cite{Pourmoghaddam2019a}.}
\label{fig:Ud_delta}
\end{figure}

Thermally pre-stressed glasses used for building applications reside only in the scale from -2 to -1 for $\log_{10}(\lambda)$ in Figure~\ref{fig:Ud_lambda}, which correspond to a range of 3 to 12 mm for $\delta$ in Figure~\ref{fig:Ud_delta}. Statistical evaluation of that relation  yields $R^2=0.93$ given the data in \cite{Pourmoghaddam2019a} and hence delivers a suitable and simple model for further analysis within the context of this paper.

Having established relations between the per-stress level of the glass pane and the characteristic size of the mean glass fracture particle, the next section is concerned with the computation of the expansion of thermally pre-stressed glass fragments.

\vspace{5mm}

\subsection{Expansion of thermally Pre-Stressed Glass Fragments}
The release of residual stresses in tempered glass leads to an overall in-plane expansion of a glass plate due to straining of the individual fragments. In \cite{Nielsen2016} it is shown how a single fragment is deforming when the pre-stress is released. 

\begin{figure}[ht!]
    \centering
    \includegraphics{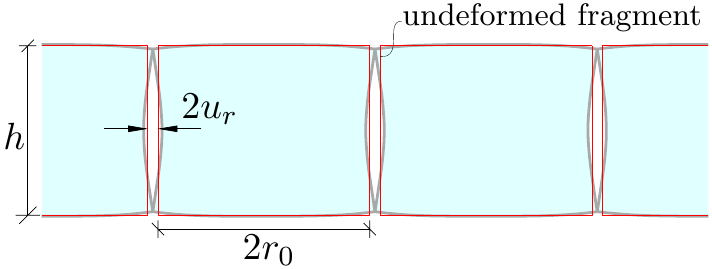}
    \caption{Expansion of a glass plate due to deformations in neighbouring fragments. The red lines indicates the undeformed fragment geometry.}
    \label{fig:MultipleFragments}
\end{figure}

The expansion of a tempered glass plate due to fragmentation is investigated by searching for the deformations of an average fragment and then integrate the contributions over a specific glass plate to obtain the total expansion as indicated in Figure~\ref{fig:MultipleFragments}.

\begin{figure}[ht!]
\includegraphics[width=.95\columnwidth]{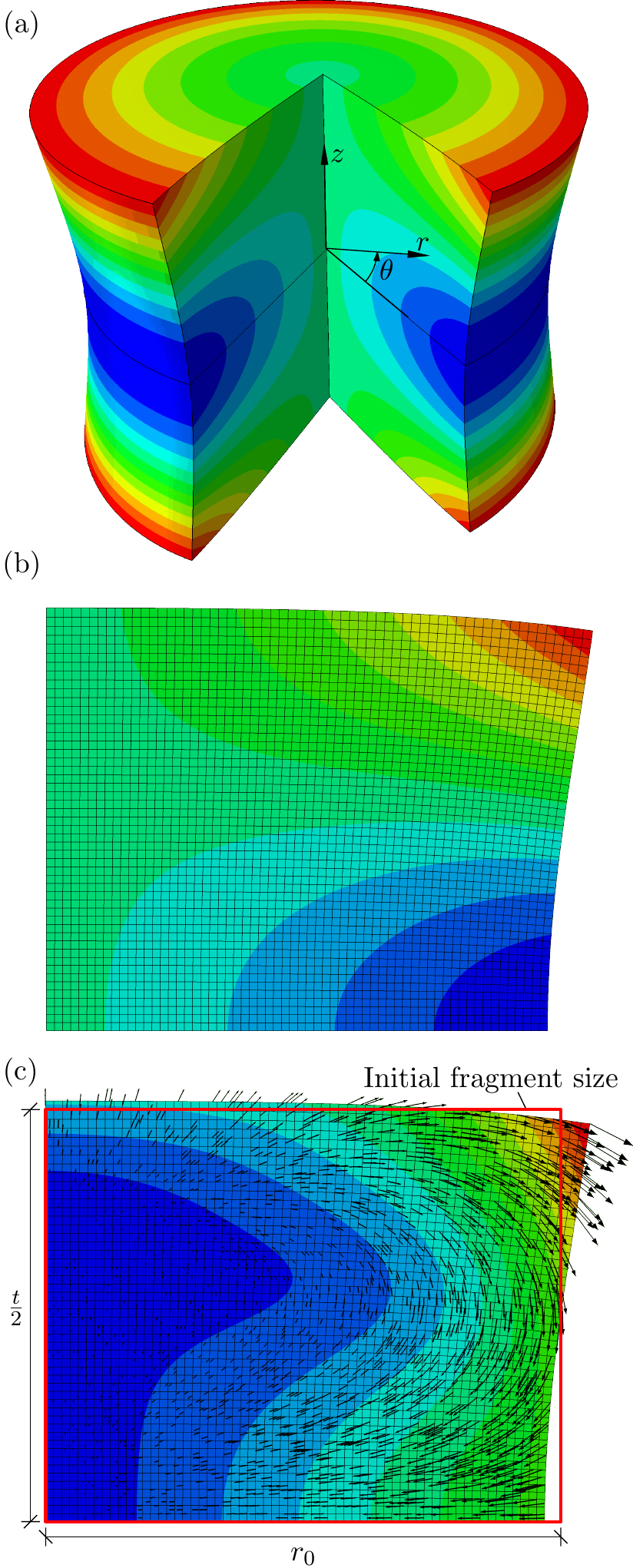}
\caption{Deformations in a fragment ($\sigma_\text{s}=\si{-75~}{MPa}$, $h=\si{8~}{mm}$, $r_0=\si{5~}{mm}$, $E=\si{70~}{GPa}$ and $\nu=\si{0.23}$). (a) Deformed cylindrical fragment. (b) Radial displacements, $u_\text{r}$, and a typical mesh. (c) Magnitude of the displacements. Arrows indicates direction and magnitude. The red square indicates the initial shape. All plots are shown with displacements magnified 100 times.}
    \label{fig:alpha1}
\end{figure}

An efficient (in terms of computational costs) axi-symmetric FE model as described in \cite{Nielsen2016} is applied. This indicates that each fragment is, initially, considered as a cylinder with height equal to the glass thickness, $h$, and radius, $r$, representing the in-plane size of the fragment Figure~\ref{fig:alpha1}.

Initially the cylinder is considered stress free, however, in the second step a parabolic stress distribution (over the height) is applied by means of a prescribed temperature field.
In the third step boundary conditions representing the neighbouring glass are removed which will represent the fragmentation of the glass. However, due to linearity (linear elastic material, small displacements and deformations) we can skip some of the intermediate steps and apply the stress state directly on the cylinder without boundary conditions (except for those needed to prevent rigid body motions). Furthermore, symmetry can be utilised in order to reduce the computational costs even further and only 1/4 of the cylindrical cross-section was meshed as indicated in Figure~\ref{fig:alpha1}. Due to the high stress variation in the fragment, a dense finite element mesh is required. For these calculations second order displacement elements were used with at least\footnote{In the FE analysis, several different models were used as the geometry (fragments size) was one of the key parameters investigated.} 100 elements through the (full height) of the fragment. This model is extremely efficient and for the results presented in this paper more than 20000 computations with varying parameters were carried out. The principles of the FE-model was experimentally validated in the paper \cite{Nielsen2017}. The output from a single simulation, as shown in Figure~\ref{fig:alpha1}, provides both stresses and deformations of a fragment. 

The primary result of interest for this study was the in-plane expansion of the fragments due to their horizontal deformation at the top and the bottom of each fragment. For this, it is assumed, that the stress distribution is homogeneous throughout the glass plate, which is a fairly good approximation from a distance of already two times the thickness from the edges. As glass plates are typically very thin compared to their width and length, the influence of the stress distribution at the edges is neglected. 

When a pre-stressed glass plate fractures, the neighbouring fragments will "push" each other, resulting in a net expansion of the plate, as sketched in Figure~\ref{fig:MultipleFragments}, which is used for calculating the total free expansion.

From the FEM-model, as shown in Figure~\ref{fig:alpha1}, a relative maximum in-plane expansion of a fragment can be found by simply dividing the maximum in-plane displacement of the fragment, $u_\text{r}$, by the size of the (undeformed) fragment, $r_0$:
\begin{equation}
    \varepsilon_\text{r} = \frac{u_\text{r}}{r_0}.
\end{equation}
This quantity is comparable to a strain and is denoted, $\varepsilon_r$, and referred to as the maximum radial strain.

By assuming that all fragments in a glass plate can be represented (on average) by the fragment used in the model, it is then possible to estimate the total free in-plane expansion of a thermally tempered glass plate. However, since we assume all fragments to expand equally, the total free expansion in any in-plane direction, $\Delta u_i$, can be calculated by simply multiplying the given plate dimension with the maximum radial strain for a representative fragment:
\begin{equation}\label{eq:ui_epsr}
    \Delta u_i = \ell_i\cdot \varepsilon_\text{r}
\end{equation}
where subscript $i$ indicates a direction and $\ell_i$ is the in-plane dimension in the $i$'th direction.

\begin{figure}[ht!]
    \centering
    \includegraphics{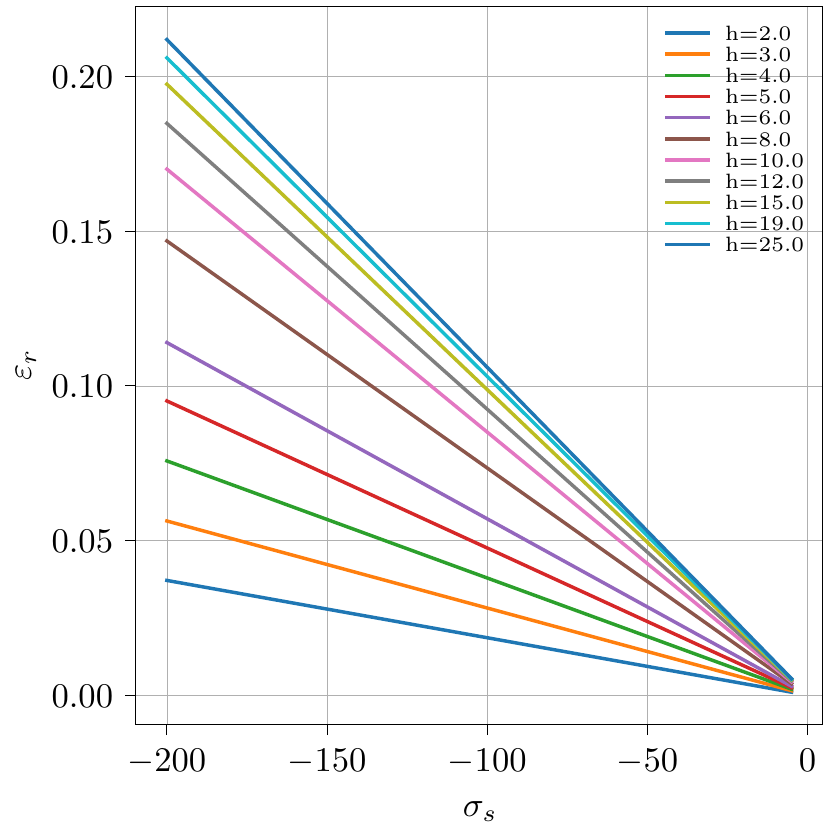}
    \caption{Radial strain of a fragment as function of the residual surface stress for different thicknesses of the glass and a fixed fragment size ($r_0=\SI{5}{\mm}$)}
    \label{fig:Epsr-vs-Sigma}
\end{figure}

From the parametric study $u_\text{r}$ is recalculated to $\varepsilon_\text{r}$ using Eq.~\eqref{eq:ui_epsr}. A plot showing this strain as a function of the surface residual stress, $\sigma_\text{s}$, for different thicknesses can be seen in Figure~\ref{fig:Epsr-vs-Sigma}. 

\begin{figure}[ht!]
    \centering
    \includegraphics{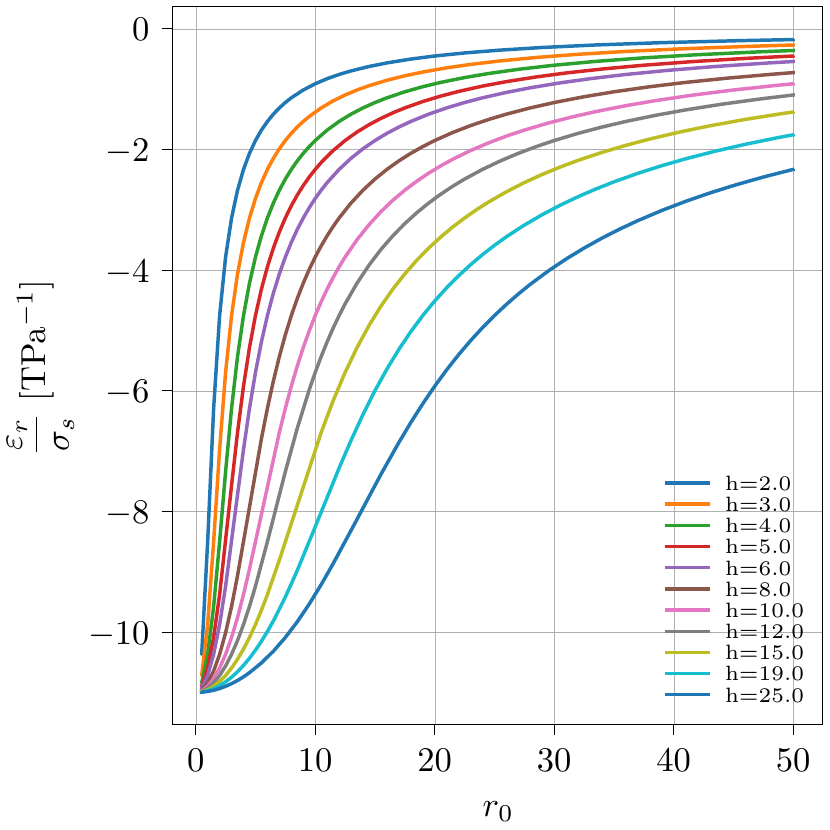}
    \caption{Nomalised radial strain versus fragment size for different thicknesses.}
    \label{fig:Eps-vs-r}
\end{figure}

From the Figure it is seen that the response is linear and it is therefore reasonable to normalise the strain with $\sigma_\text{s}$. Doing this allows us to plot the variations with the fragment size, $r_0$. This is shown in Figure~\ref{fig:Eps-vs-r}.

\begin{figure}[ht!]
    \centering
    \includegraphics{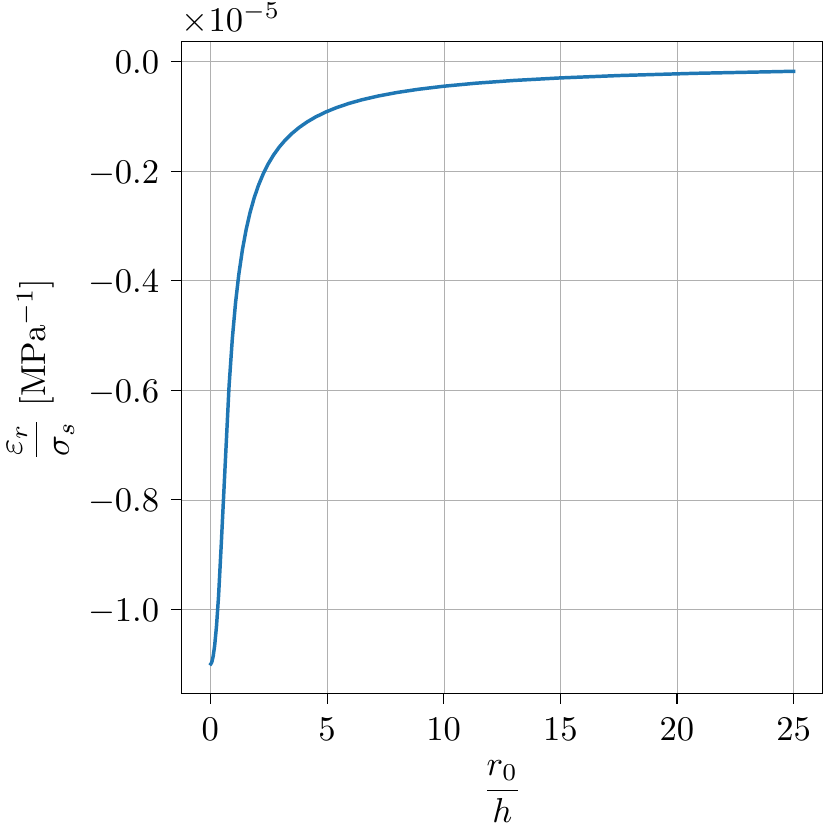}
    \caption{Normalised radial strain versus normalised fragment size.}
    \label{fig:Eps-vs-rh}
\end{figure}

From the Figure~\ref{fig:Eps-vs-r} it can be noticed that all curves seems to have the same overall shape and normalising the horizontal axis with the thickness, $h$, yields all curves to coincide. This is shown in Figure \ref{fig:Eps-vs-rh}.

\section{Equivalent Temperature Difference (ETD) Model for in-plane Expansion of Fractured Thermally Pre-stressed Glass Panes} \label{sec_ETD}

Modelling concrete shrinkage effects and primary as well as secondary effects on the composite structure via equivalent temperature differences (ETD) and induced linear expansion and / or curvature is well known and established for steel-concrete composite structures in both academia and engineering practice \cite{johnson2004composite,vayas2013design}. This paper takes the foundations laid so far to elaborate a \emph{equivalent temperature differences} (ETD) model upon a thermal expansion analogy for strains to provide a handy method of estimating the average amount of in-plane expansion of thermally pre-stressed glass panes.

\subsection{Deriving the ETD model for the free expansion of tempered glass at failure}

In analogy to the definition of thermal strains via a thermal expansion coefficient, $\alpha_\text{th}$, and a governing temperature difference, $\Delta T$,
 \begin{equation}
 \label{eq:ThermalStrain}
     \varepsilon_\text{th}  =\alpha_\text{th} \Delta T 
\end{equation}
a fracture expansion coefficient, $\alpha_\text{fr}$ can be defined in order to establish a relation between the free expansion strain caused by fragmentation of the tempered glass, $\varepsilon_\text{fr}$, and the residual stress state, quantified by $\sigma_\text{s}<0$;
%
\begin{equation}\label{eq:fracStrain}
      \varepsilon_\text{fr} = \alpha_\text{fr}\frac{\nu-1}{E}\sigma_\text{s}.
\end{equation}
The term, $(\nu-1)/E$, is governing the in-plane behaviour of a plate and, $E$, and, $\nu$, represents Young's modulus and Poisson's ratio respectively. The free expansion strain caused by the fracture, $\varepsilon_\text{fr}$, can be interpreted as the total strain of a given tempered glass plate upon fragmentation.

Enforcing compatibility of the ''strain from fracture'', $\varepsilon_\text{fr}$, with the maximum radial strain, $\varepsilon_\text{r}$, delivers:
\begin{equation}
     \varepsilon_\text{r} = \varepsilon_\text{fr}.
\end{equation}

\begin{figure*}[ht!]
\begin{overpic}[]{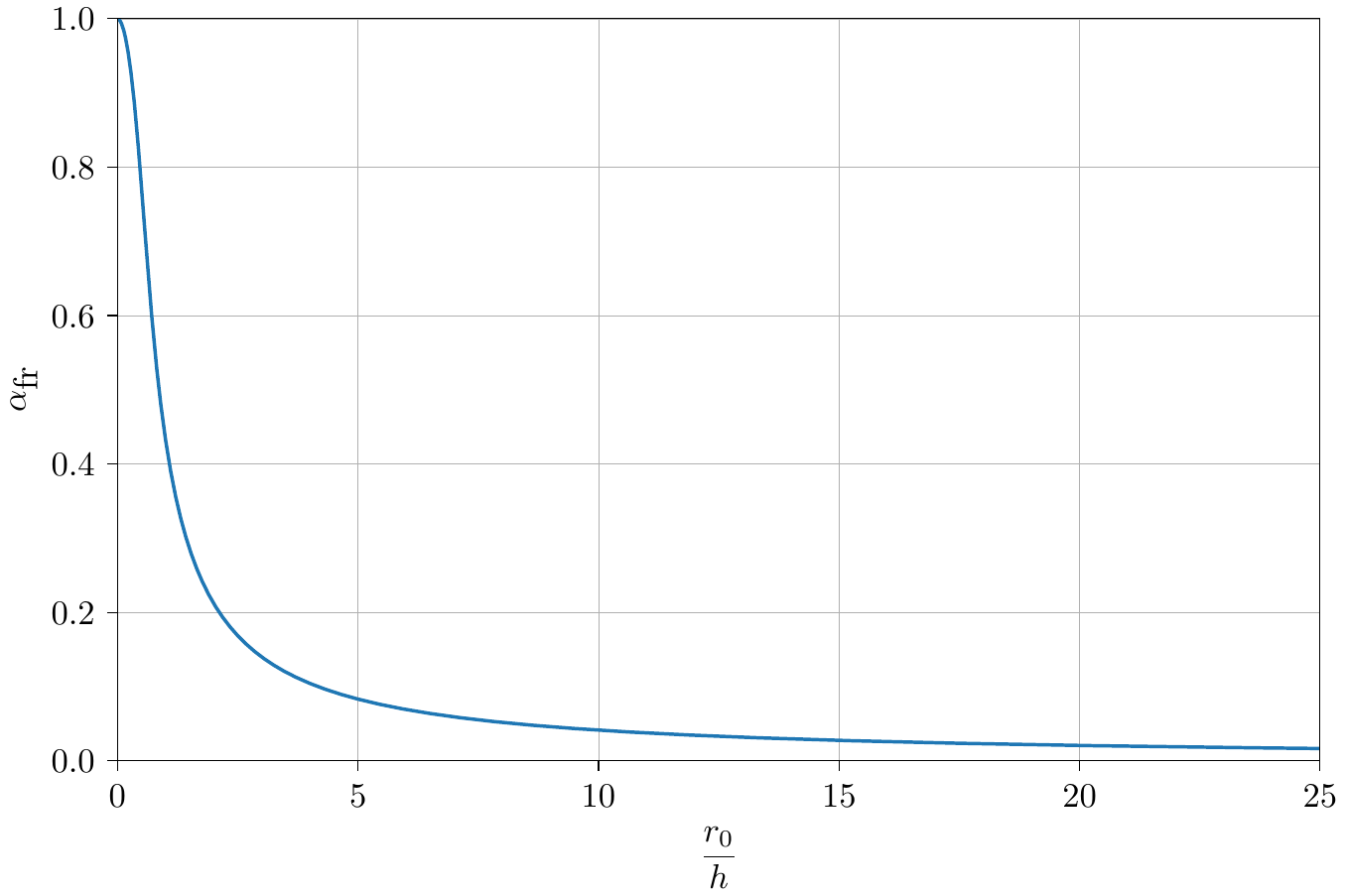}
\put(28.3,18){\frame{\includegraphics[]{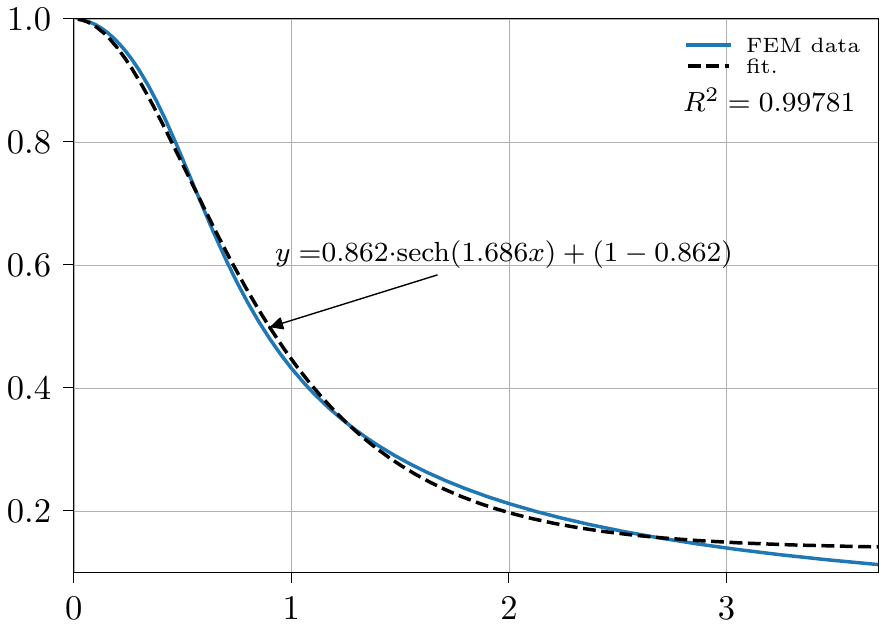}}}
\put(0.2,0){\polygon(9.0,15.5)(21.6,15.5)(21.6,65.3)(9.0,65.3)}
\put(21.8,60){\line(2,-1){6.5}}
\end{overpic}
\caption{Relation between $\alpha_\text{fr}$ and the ratio between the fragment size (radius) and the glass thickness, $r/h$. Notice the zoom providing results for more typical values of $r/h$.}
    \label{fig:alpha2}
\end{figure*}

The fracture expansion coefficient, $\alpha_\text{fr}$, from Eq.~\eqref{eq:fracStrain} can be determined through the FE-analysis on the radial strain, $\varepsilon_\text{r}$, as carried out above and summarised in the plot shown in Figure \ref{fig:Eps-vs-rh}. Multiplying the curve in Figure~\ref{fig:Eps-vs-rh} with $E/(\nu-1)$ as indicated in Eq.~\eqref{eq:fracStrain} the fracture expansion coefficient, $\alpha_\text{fr}$, can be found as a function of the fragment size relative to the glass thickness, $r_0/h$, as shown in Figure~\ref{fig:alpha2}. The upper limit for the fracture expansion coefficient is $\alpha_\text{fr}=1$, which corresponds to a zero fragment size. The physical interpretation of this is that if the glass is completely pulverised, all residual stresses are released and are all converted linear elastically into the fracture expansion strain. For (unrealistically) large fragment sizes the curve tends towards zero indicating that the relative amount of residual stresses converted into fracture expansion strain approaches zero. 

In Figure~\ref{fig:alpha2}, the first and, from a practical point of view, most relevant part of the curve is fitted using a hyperbolic secant function, $\sech{x}=\frac{1}{\cosh{x}}$, as this function have the right properties with horizontal asymptotes for $x=0$ and $x\rightarrow \infty$. A function on the form:
\begin{equation}\label{eq_alp0}
    \alpha_\text{fr}\left(\frac{r_0}{h}\right) = a_1\sech{\left(b_1\frac{r_0}{h}\right)}+\left(1-a_1\right)
\end{equation}
was therefore fitted to the plot for $\frac{r_0}{h}\leq 3.7$ with relatively good agreement as shown in Figure \ref{fig:alpha2}. This value corresponds roughly to $\sigma_\text{s}=\SI{-50}{MPa}$ for a $\SI{3}{\mm}$ glass which we will consider maximum fragment size for a standard thickness thermally pre-stressed glass.

Within this paper in the sense of a statistical first order expectation approximation it is assumed, that the radial strain shown in Figure~\ref{fig:alpha1} is fully contributing to the plate's expansion after fracture. Hence, this paper specifies the nomenclature "$\left\langle \cdot \right\rangle$" to formally emphasise, that the derived quantity is to be interpreted as a statistical first-order approximation of the expected value of the respective quantity, so that Eq.~\ref{eq:ThermalStrain} and ~\ref{eq:fracStrain} yield: 

\begin{equation}
 \label{eq:ThermalAnalogyStrainOfFracture}
\left\langle \varepsilon_\text{th} \right\rangle = \left\langle \varepsilon_\text{fr} \right\rangle \hspace{.2cm} \iff \hspace{.2cm} \alpha_\text{th} \left\langle \Delta T \right\rangle = \alpha_\text{fr}\frac{\nu-1}{E}\sigma_\text{s}
\end{equation}

which after rearranging yields:
\begin{equation}
\label{eq:ThermalAnalogyStrainOfFracture2}
\left\langle \Delta T \right\rangle = \frac{\alpha_\text{fr}}{ \alpha_\text{th} } \frac{\nu-1}{E}\sigma_\text{s}
\end{equation}

Using the proposed ETD method allows the use of both analytical and commercial software to compute an estimate of the effects of the stress state in adjacent structural elements, see e.g. the examples in Section~\ref{sec:example2} .

\section{Summary and Application Examples}



In the previous section a model relating the free expansion of a tempered glass plate with the residual surface stresses, $\sigma_\text{s}$, was derived. In this section we will summarise and provide some examples of usage, repeating some of the key equations in the model for the convenience of the reader.

The fracture expansion strain can be calculated from Eq.~\eqref{eq:fracStrain}:
\begin{equation}\label{eq_exp}
    \varepsilon_\text{fr} = \alpha_\text{fr}\frac{\nu-1}{E}\sigma_\text{s} \quad \Leftrightarrow \quad \alpha_\text{fr}=\frac{\varepsilon_\text{fr}}{\sigma_\text{s}}\frac{E}{\nu-1}
\end{equation}
In this equation Young's modulus, $E$, and Poisson's ratio, $\nu$, for glass can be found in Table \ref{tab:Parameters}. The fracture expansion coefficient, $\alpha_\text{fr}$, can be estimated from the FEM study reported in Figure \ref{fig:alpha2} in which the most relevant part for tempered glass is fitted by an expression in form of:
\begin{equation}\label{eq_alp}
    \alpha_\text{fr}\left(\frac{r_0}{h}\right) = a_1\sech{\left(b_1\frac{r_0}{h}\right)}+\left(1-a_1\right)
\end{equation}
where the constants $a_1$ and $b_1$ can be found in Table \ref{tab:Parameters}. The thickness of the glass, $h$, is assumed known. 

The mean fragment size, $r_0$, can be estimated from \cite{Pourmoghaddam2019a} (cf. Eq.~ \eqref{eq_r_0_simple}) and the relation $\delta=2r_0$ as:
\begin{equation}\label{eq_r0}
    r_0 = \frac{a_2}{U_\text{D}}
\end{equation}

where, $a_2$, can be found in Table \ref{tab:Parameters} and, $U_\text{D}$, is the strain energy density for tempered glass given by Eq.~\eqref{eq:U_04}:
\begin{equation}\label{eq_UD}
    U_\text{D} = \frac{1-\nu}{5E}\sigma_\text{s}^2
\end{equation}

\begin{table}[H]
\caption{Parameters to be used for the ETD model.}\label{tab:Parameters}
\begin{tabular}{c | c}
Parameter               & Reference  \\ \hline
$E$=\SI{70}{\GPa}         &                        \\
$\nu$=\num{0.23}          &                        \\ 
$\alpha_\text{th}=\SI{9.1e-6}{\per\kelvin}$    &                        \\ \hline
$a_1$=\num{0.862}        & \multirow{2}{*}{Figure \ref{fig:alpha2}}\\
$b_1$ = \num{1.686}      &                                \\ \hline
$a_2$ = \SI{61.05}{\joule\per\meter\squared}   &  Eq.~\eqref{eq_r_0_simple}      \\ \hline                      
\end{tabular}
\end{table}

Combining Eqs.~\eqref{eq_exp},\eqref{eq_alp}, \eqref{eq_r0} and Eq.~\eqref{eq_UD} we obtain:
\begin{equation}
\begin{split}
\varepsilon_\text{fr} =&
      \frac{\nu-1}{E}\left(a_1\sech{\left(\frac{5 E a_{2} b_{1}}{h \sigma_\text{s}^{2} \left(1 - \nu\right)} \right)}+1-a_1\right)\sigma_\text{s}
\end{split}
\end{equation}

Inserting all constants from Table \ref{tab:Parameters} and rearranging, the expression can be written as:
\begin{equation}\label{eq:eps_simple}
\begin{split}
\varepsilon_\text{fr} &=
\frac{-\sigma_\text{s}}{\SI{658.8e3}{\MPa}}\left[1\phantom{6.246\sech{\left(\frac{\SI{4.679e16}{\meter\MPa\squared}}{h\sigma_s^2}\right)}}\right.\\
&\left.+6.246\sech{\left(\frac{\SI{4.679e16}{\meter\MPa\squared}}{h\sigma_s^2}\right)}\right]
\end{split}
\end{equation}

From Eq.~\eqref{eq:eps_simple} a plot showing the fracture strain, $\varepsilon_\text{fr}$, as a function of the residual surface stress, $\sigma_\text{s}$, for different glass thicknesses, $h$, is generated and shown in Figure \ref{fig:Epsfr-sig-h}:

\begin{figure*}[ht!]
    \centering
    \includegraphics{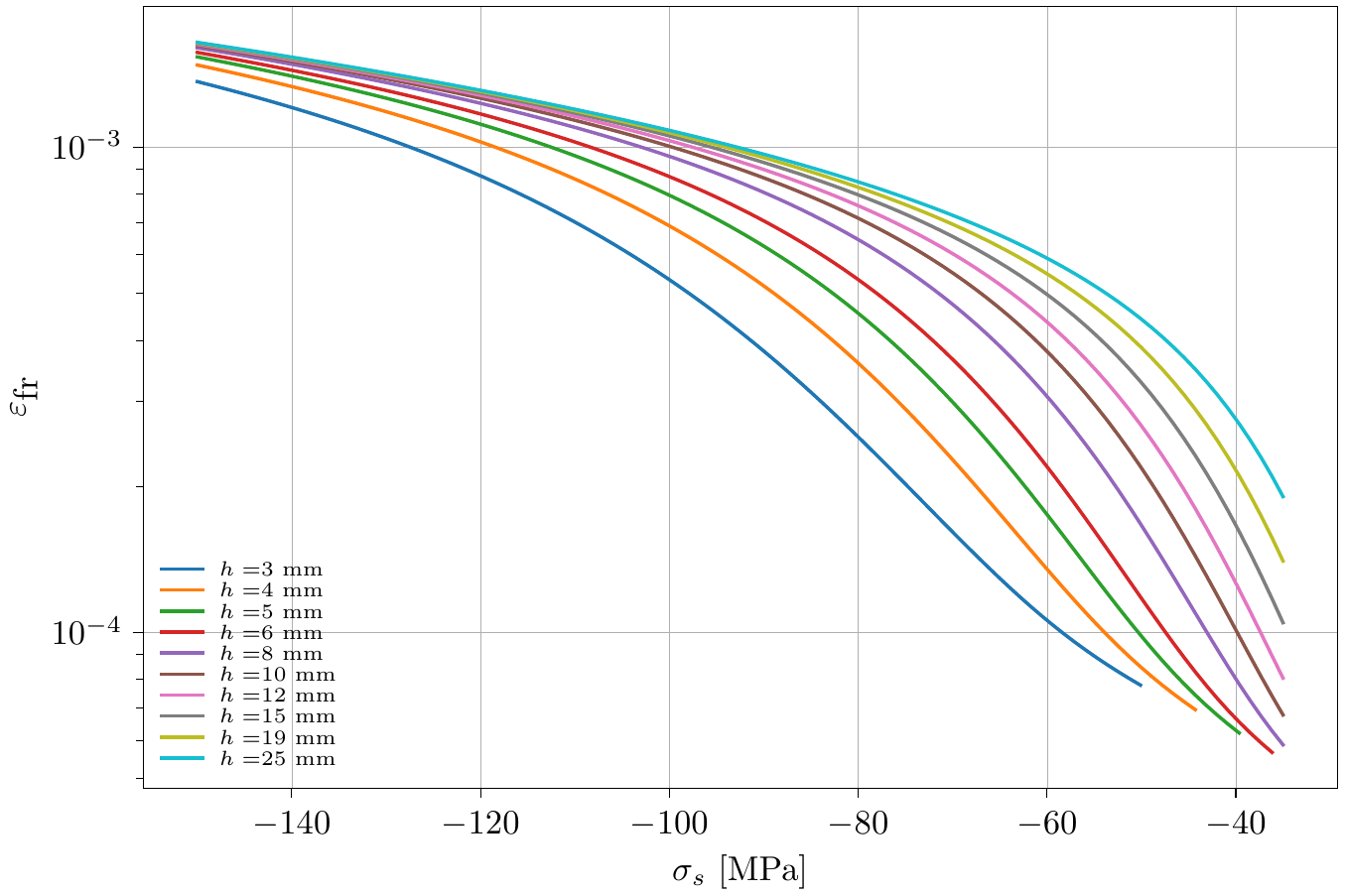}
    \caption{The fracture strain as a function of the residual surface compression for different glass thicknesses.}
    \label{fig:Epsfr-sig-h}
\end{figure*}

\subsection{\emph{Example 1: ETD Model Prediction for the expansion due to fracture of a mono tempered glass pane}}\label{sec:example1}

In this example we will consider a $\ell_x \times \ell_y = \SI{2.5x1}{\m}$, \SI{10}{\mm} thick monolithic tempered glass pane. The residual surface stress is measured to $\sigma_s = \SI{-85}{MPa}$. The total free expansion in case of failure can now be estimated using Eq.~\eqref{eq:ui_epsr} and Eq.~\eqref{eq:eps_simple} or Figure \ref{fig:Epsfr-sig-h}.

According to Eq.~\eqref{eq:eps_simple}, the fracture strain in the glass pane is $\varepsilon_\text{fr}=\num{791e-6}$. The total expansion in the $x$ and $y$ directions, then becomes:
\begin{equation}
\begin{split}
    u_x &= \varepsilon_\text{fr}\cdot\ell_x =\num{791e-6}\cdot\SI{2500}{\mm}= \SI{1.98}{\mm}\\ 
    u_y &= \varepsilon_\text{fr}\cdot\ell_x =\num{791e-6}\cdot\SI{1000}{\mm}=\SI{0.791}{\mm} \\ 
\end{split}
\end{equation}

The equivalent temperature for use inside a Finite Element software can be computed using Eq. \ref{eq:ThermalAnalogyStrainOfFracture2}, which for this example yields:
\begin{equation}
\label{eq:ThermalAnalogyStrainOfFracture_example1}
\left\langle \Delta T \right\rangle = 88 K 
\end{equation}
assuming $\alpha_\text{th}=\SI{9.1e-6}{\per\kelvin}$ which is commonly used for glass.

\subsection{\emph{Example 2: ETD model prediction for partly fractured laminated glass}}\label{sec:example2}
The model can also be used for estimating the extra load on intact panes in (partly) fractured laminated glass.

\begin{figure}[ht!]
    \centering
    \includegraphics{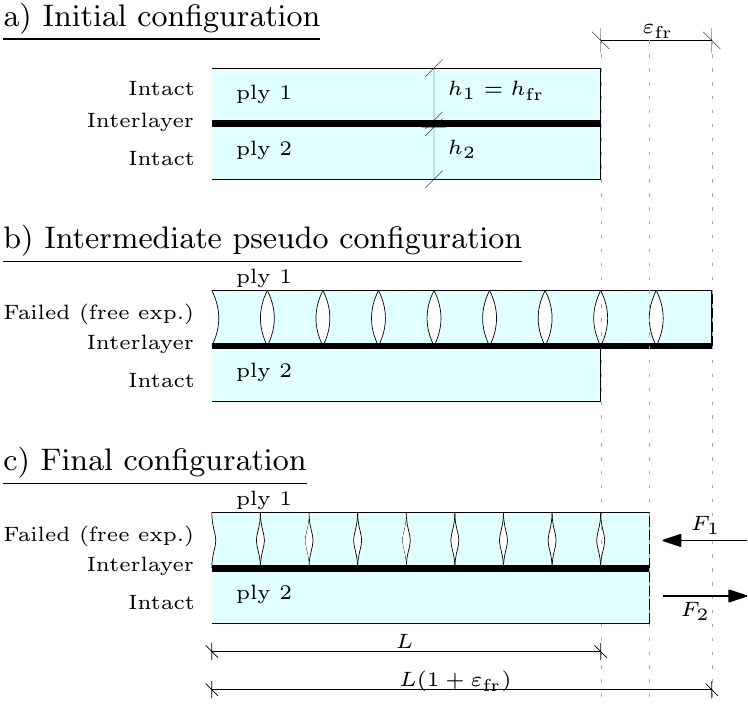}
    \caption{Laminated glass, ply 1 is tempered glass. a) shows the initial configuration, b) shows the free expansion of the tempered glass and c) shows the total elongation of the laminated glass.}
    \label{fig:Example_laminate01}
\end{figure}

Considering a two ply laminated glass as shown in Figure~\ref{fig:Example_laminate01}a. In the initial configuration, no external stresses are present and the two plies have the thickness $h_1$ and $h_2$. 

In the intermediate pseudo configuration, see Figure~\ref{fig:Example_laminate01}b, the fractured tempered glass (ply 1) is shown with its free expansion and the intact glass (ply 2) is not affected by this. Obviously, the expansion will be transferred through the interlayer and a first approximation to the final state can be found by assuming a stiff interlayer and no rotations (bending), see Figure~\ref{fig:Example_laminate01}c. Now we can find the forces in the final configuration by first equal length for the two plies:
\begin{equation}
\begin{split}
        L(1+\varepsilon_\text{fr})+L(1+\varepsilon_\text{fr})\cdot\varepsilon_1 &= L+L\cdot\varepsilon_2 \qquad \Leftrightarrow\\
        \varepsilon_\text{fr}+(1+\varepsilon_\text{fr})\cdot\varepsilon_1 &= \varepsilon_2\\
\end{split}
\end{equation}

Now applying Hookes law and requiring equilibrium, $F_1=-F_2=F$ we find:
\begin{equation}
        \varepsilon_\text{fr}+(1+\varepsilon_\text{fr})\cdot\frac{-F}{E_1h_1} - \frac{F}{E_2h_2}=0
\end{equation}

from which the force, $F$, and the stress in the intact ply can be found as:
\begin{equation}\label{eq:example02}
    F=h_2\sigma_2=\frac{h_1E_1h_2E_2\varepsilon_{\text{fr}}}{h_1E_1+(1+\varepsilon_\text{fr})h_2E_2}.
\end{equation}

As an example, one could consider the \SI{10}{\mm} glass with a surface residual stress, $\sigma_{\text{s}}=\SI{-85}{MPa}$ from Section~\ref{sec:example1} for both layers in a two-ply laminated glass plate. If the one layer fails, the stresses in the intact layer can be estimated from Eq.~\eqref{eq:example02}. Assuming same stiffness for both plies, $h_1=h_2=\SI{10}{\mm}$ and $E_1=E_2=\SI{70}{\GPa}$ we find, in this case, the stress in the intact pane to be $\sigma=\SI{28}{\MPa}$.

An unknown in the model is the compressive stiffness for fractured tempered glass. However, to the best knowledge of the authors, such investigations have not yet been published. It is expected that the stiffness of the broken glass is lower compared to the intact glass and is likely to be strain dependent.

\begin{figure}[ht!]
    \centering
    \includegraphics{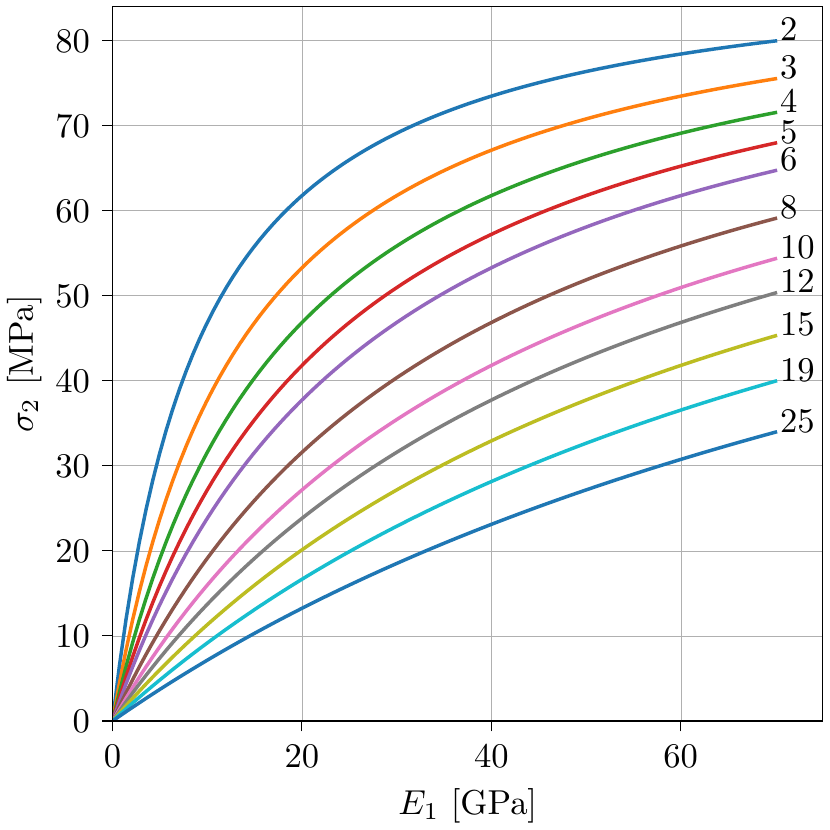}
    \caption{Peak stress in intact ply, $\sigma_2$, as a function of the stiffness in the broken ply, $E_1$, for different thicknesses of the intact ply, $h_2$ (in \si{mm}). The broken ply is having a thickness of $h=\SI{15}{\mm}$ and a residual surface stress of $\sigma_\text{s}=\SI{-120}{\MPa}$.}
    \label{fig:sig-intact-stiffness}
\end{figure}

The plot in Figure~\ref{fig:sig-intact-stiffness} shows an example of the tensile stress in the intact layer, $\sigma_2$, as a function of the compressive stiffness of the broken layer, $E_1$, for varying thicknesses of the intact layer, $h_2$. From the results in the figure, it is seen that combining a \SI{15}{\mm} fully tempered glass with e.g. a \SI{6}{\mm} thick annealed glass may cause problems in case of failure of the tempered glass as the peak stress in the annealed glass may reach up to \SI{65}{\MPa}.

The proposed model does not take into account the stiffness of the inter-layer and bending of the plies due to asymmetric failure. However, the fracture velocity of tempered glass has been measured using high-speed cameras to be approximately \SI{1466}{\meter\per\second} \cite{Nielsen2009}, indicating a very high loading rate of the inter-layer and thereby a dynamic problem. It is well known, that common interlayers, such as Poly Vinyl Butyral (PVB), Ethylene Vinyl Acetate (EVA) and ionomers e.g. SentryGlas\textsuperscript{\textregistered}, shows an increase in stiffness with the loading rate. This supports the assumption of a full shear transfer between the plies, however, it also indicates that some of the load might actually be carried by the interlayer, which is not accounted for in the current model. Due to the dynamic nature of the problem, the strength of the intact glass may also be higher than what is often assumed for quasi static problems, see e.g. \cite{Meyland2021} where a review on available strength data for soda-lime-silica glass is given.




\section{Summary and Conclusion}
This paper first presented experimental and theoretical background on the glass fracture process for thermally pre-stressed glasses. The 2D-macro-scale fragmentation of glass can be basically described by the fragment particle size, $\delta$, and the fracture particle intensity, $\lambda$, which both are related to the pre-stress induced strain energy density, $U_\text{D}$, before fracture. Then further details on Finite Element (FE) simulations of single cylindrical glass particles are reported, which allowed to establish functional relations of the glass fragment particle dimensions, the pre-stress level and the resulting maximum in-plane deformation. These results are then combined with the two-parameter fracture pattern modelling to furnish an \emph{equivalent temperature differences} (ETD) for describing the in-plane expansion of thermally pre-stressed glass panes due to fracturing. Finally, two examples from engineering practice demonstrated the application of the developed graphs and formulas for further use in analytical as well as FE analysis of fractured glass laminates.

Further analysis of the fracture particle statistics vs. strain energy density proved existence of two glass fracture statistic regions. Despite that novel finding within Section~\ref{subsec:FractStats}, for engineering practice it is sufficient to only concentrate on one of the two fracture domains (\SI{3}{mm} to \SI{12}{mm} fragment particle size). The idea of ETD then was applicable straight forward for relating FE analysis results of the deformations of a single fragment upon failure of the glass with the total free expansion of a piece of tempered glass from its initial dimensions and residual surface compressive stress by using a thermal strain analogy. Here, the derived ETD model provides a tool for estimating the free (unconstrained) expansion of tempered glass at failure, which is not possible at all at the moment. To that end, our approach in form of handy ETD load cases within both, analytical as well as FE analysis, allows for the estimation of (i) secondary effects in the fractured laminate such as fracture-expansion-induced deformations or stresses, and (ii) effects due to second order influences on residual load bearing capacity of the fractured glass laminate as well as of support structures or remaining parts of glass laminates. Furthermore, a simple analytical model for estimating the peak stress in the intact ply in a partly broken laminated glass plate is provided. From the model it is found that the peak stress must be considered relevant and the model also suggest that care should be taken if laminating glass plies with too different thicknesses or mixing both tempered and annealed glass. Future research needs to address experimental validation of this ETD model on various glass laminates, where level of pre-stress as well as laminate size and glass thicknesses are varied. Furthermore, this ETD model needs to be enhances for the influence of different interlayer types as these possess pronounced differences in stiffness and hence shear coupling of the glass panes during fracture and speed of redistribution of internal forces in the post-fractured state.



\bibliographystyle{BibTeX/spbasic_unsort} 

\bibliography{01main.bib}          



\end{document}